\documentclass[useAMS,usenatbib]{mnras}
\usepackage{amsmath}
\usepackage{amssymb}
\usepackage{graphicx}
\newcommand\textlcsc[1]{\textsc{\MakeLowercase{#1}}}

\newcommand{\bff}{\bf}
\renewcommand{\bff}{}   % To remove all bold face
\def\jref@jnl#1{{\rm#1}}

\def\aj{\jref@jnl{AJ}}                   % Astronomical Journal
\def\araa{\jref@jnl{ARA\&A}}             % Annual Review of Astron and Astrophys
\def\apj{\jref@jnl{ApJ}}                 % Astrophysical Journal
\def\apjl{\jref@jnl{ApJ}}                % Astrophysical Journal, Letters
\def\apjs{\jref@jnl{ApJS}}               % Astrophysical Journal, Supplement
\def\ao{\jref@jnl{Appl.~Opt.}}           % Applied Optics
\def\apss{\jref@jnl{Ap\&SS}}             % Astrophysics and Space Science
\def\aap{\jref@jnl{A\&A}}                % Astronomy and Astrophysics
\def\aapr{\jref@jnl{A\&A~Rev.}}          % Astronomy and Astrophysics Reviews
\def\aaps{\jref@jnl{A\&AS}}              % Astronomy and Astrophysics, Supplement
\def\azh{\jref@jnl{AZh}}                 % Astronomicheskii Zhurnal
\def\baas{\jref@jnl{BAAS}}               % Bulletin of the AAS
\def\jrasc{\jref@jnl{JRASC}}             % Journal of the RAS of Canada
\def\memras{\jref@jnl{MmRAS}}            % Memoirs of the RAS
\def\mnras{\jref@jnl{MNRAS}}             % Monthly Notices of the RAS
\def\pasa{\jref@jnl{PASA}}
\def\pra{\jref@jnl{Phys.~Rev.~A}}        % Physical Review A: General Physics
\def\prb{\jref@jnl{Phys.~Rev.~B}}        % Physical Review B: Solid State
\def\prc{\jref@jnl{Phys.~Rev.~C}}        % Physical Review C
\def\prd{\jref@jnl{Phys.~Rev.~D}}        % Physical Review D
\def\pre{\jref@jnl{Phys.~Rev.~E}}        % Physical Review E
\def\prl{\jref@jnl{Phys.~Rev.~Lett.}}    % Physical Review Letters
\def\pasp{\jref@jnl{PASP}}               % Publications of the ASP
\def\pasj{\jref@jnl{PASJ}}               % Publications of the ASJ
\def\qjras{\jref@jnl{QJRAS}}             % Quarterly Journal of the RAS
\def\skytel{\jref@jnl{S\&T}}             % Sky and Telescope
\def\solphys{\jref@jnl{Sol.~Phys.}}      % Solar Physics
\def\sovast{\jref@jnl{Soviet~Ast.}}      % Soviet Astronomy
\def\ssr{\jref@jnl{Space~Sci.~Rev.}}     % Space Science Reviews
\def\zap{\jref@jnl{ZAp}}                 % Zeitschrift fuer Astrophysik
\def\nat{\jref@jnl{Nature}}              % Nature
\def\iaucirc{\jref@jnl{IAU~Circ.}}       % IAU Cirulars
\def\aplett{\jref@jnl{Astrophys.~Lett.}} % Astrophysics Letters
\def\apspr{\jref@jnl{Astrophys.~Space~Phys.~Res.}}
\def\bain{\jref@jnl{Bull.~Astron.~Inst.~Netherlands}} 
\def\fcp{\jref@jnl{Fund.~Cosmic~Phys.}}  % Fundamental Cosmic Physics
\def\gca{\jref@jnl{Geochim.~Cosmochim.~Acta}}   % Geochimica Cosmochimica Acta
\def\grl{\jref@jnl{Geophys.~Res.~Lett.}} % Geophysics Research Letters
\def\jcp{\jref@jnl{J.~Chem.~Phys.}}      % Journal of Chemical Physics
\def\jgr{\jref@jnl{J.~Geophys.~Res.}}    % Journal of Geophysics Research
\def\jqsrt{\jref@jnl{J.~Quant.~Spec.~Radiat.~Transf.}}
\def\memsai{\jref@jnl{Mem.~Soc.~Astron.~Italiana}}
\def\nphysa{\jref@jnl{Nucl.~Phys.~A}}   % Nuclear Physics A
\def\physrep{\jref@jnl{Phys.~Rep.}}   % Physics Reports
\def\physscr{\jref@jnl{Phys.~Scr}}   % Physica Scripta
\def\planss{\jref@jnl{Planet.~Space~Sci.}}   % Planetary Space Science
\def\procspie{\jref@jnl{Proc.~SPIE}}   % Proceedings of the SPIE

\title[Peculiar velocity and metallicity patterns]{Spiral-induced velocity and metallicity patterns in a cosmological zoom simulation of a Milky Way-sized galaxy}
\author[Grand et al.]{\parbox[t]{\textwidth}{
Robert J. J. Grand$^{1,2}$\thanks{robert.grand@h-its.org}, Volker Springel$^{1,2}$, Daisuke Kawata$^3$, Ivan Minchev$^4$, Patricia S\'{a}nchez-Bl\'{a}zquez$^{5,6}$, Facundo A. G\'{o}mez$^7$, Federico Marinacci$^8$, R\"{u}diger Pakmor$^1$ and David J. R. Campbell$^9$} \vspace{10pt} \\
$^1$Heidelberger Institut f\"{u}r Theoretische Studien, Schloss-Wolfsbrunnenweg 35, 69118 Heidelberg, Germany\\
$^2$Zentrum f\"{u}r Astronomie der Universit\"{a}t Heidelberg, Astronomisches Recheninstitut, M\"{o}nchhofstr. 12-14, 69120 Heidelberg, Germany\\
$^3$Mullard Space Science Laboratory, University College London, Holmbury St. Mary, Dorking, Surrey, RH5 6NT, United Kingdom \\
$^4$Leibniz-Institut f\"{u}r Astrophysik Potsdam (AIP), An der Sternwarte 16, D-14482, Potsdam, Germany \\
$^5$Departamento de F\'{i}sica Te\'{o}rica, Universidad Aut\'{o}noma de Madrid, Cantoblanco, E28049, Spain \\
$^6$Instituto de Astrof\'{i}sica, Universidad Pontifica Cat\'{o}lica de Chile, Av. Vicu\~{n}a Mackenna 4860, Santiago, Chile \\
$^7$Max-Planck-Institut f\"{u}r Astrophysik, Karl-Schwarzschild-Str. 1, D-85748, Garching, Germany  \\
$^8$Department of Physics, Kavli Institute for Astrophysics and Space Research, MIT, Cambridge, MA 02139, USA \\
$^9$Institute for Computational Cosmology, Department of Physics, Durham University, South Road, Durham, DH1 3LE, UK\\
}

\date{Accepted XXX. Received YYY; in original form ZZZ}

\pubyear{2015}
    
\begin{document}

\label{firstpage}
\label{lastpage}
\pagerange{\pageref{firstpage}--\pageref{lastpage}}
\maketitle

\begin{abstract}
We use a high resolution cosmological zoom simulation of a Milky Way-sized halo to study the observable features in velocity and metallicity space associated with the dynamical influence of spiral arms. For the first time, we demonstrate that spiral arms, that form in a disc in a fully cosmological environment with realistic galaxy formation physics, drive large-scale systematic streaming motions. In particular, on the trailing edge of the spiral arms the peculiar galacto-centric radial and {\bff azimuthal} velocity field is directed radially outward and {\bff azimuthal}ly backward, whereas it is radially inward and {\bff azimuthal}ly forward on the leading edge. Owing to the negative radial metallicity gradient, this systematic motion drives, at a given radius, an azimuthal variation in the residual metallicity that is characterised by a metal rich trailing edge and a metal poor leading edge. We show that these signatures are  theoretically observable in external galaxies with Integral Field Unit instruments such as VLT/MUSE, and if detected, would provide evidence for large-scale systematic radial migration driven by spiral arms.
\end{abstract}


\section{Introduction}

Spiral arms are typical features of late-type disc galaxies, which make up roughly $\sim 70 \%$ of the bright galaxies in the local volume. They are found not only in the distribution of cold gas and young, bright stars, but also in the old stellar populations \citep{RZ95}, which indicate that they are a dynamical phenomenon. Most theoretical models in the last 50 years have centred on variations of the classic density wave theory \citep{LS64} or swing amplification theory \citep{JT66}. In the former, the spiral density enhancement is regarded as a crest of stars that preserves its shape as it propagates around the disc, whereas in the latter it is described by a shearing over-density that grows and decays around a preferred pitch angle \citep[e.g.,][]{BSW12,GKC13,MK14}. In the last decades many observational attempts have been made to test these theories, using methods such as the Tremaine-Weinberg equations \citep{TW84} and the spatial distribution of star forming tracers \citep[e.g.,][]{FR10,FCK12}. However, despite decades of study, the nature of spiral arms remains an unsolved problem in contemporary astrophysics.

In recent years, numerical simulations have provided new insights into the formation and evolution of spiral arms. Crucially, $N$-body simulations commonly show transient spiral arms \citep{Se11}, which form and disrupt on the order of a dynamical time. An important secular effect of transient spiral arms is their ability to change the angular momentum of individual star particles around the co-rotation radius, which over several dynamical times leads to significant radial mixing of star particles throughout the disc \citep{SB02}. \citet{GKC11} showed that transient spiral arms appear to wind-up because of their co-rotation with disc particles at all radii, which can induce systematic motion of star particles along the leading and trailing edges of the spiral arms \citep{GKC13b,KHG14}, in addition to changing the metal distribution in the disc \citep{GKC15}.

The majority of simulations used for the study of spiral arms are set up in isolated, idealised initial conditions, such as pure exponential discs set up in equilibrium and in the absence of many important aspects of galaxy formation physics such as stellar accretion and gas inflows/outflows. While a major advantage of these simulations is the ability to control parameters and study individual effects, the lack of a realistic cosmological setting can limit their predictive power in the sense that the predicted relations between kinematics, metallicity and age may be somewhat artificial, particularly over cosmological time-scales. In this Letter, we show for the first time systematic radial migration around the spiral arms in a high-resolution self-consistent cosmological simulation which includes a wide range of galaxy formation physics. This implies that the nature of spiral arms is similar to the transient,  winding spiral density enhancements commonly seen in idealised $N$-body simulations. Furthermore, we demonstrate for the first time that the systematic motion drives clear patterns of azimuthal variation of the metallicity distribution. We demonstrate that the peculiar velocity field and metallicity distribution around the spiral arms can be detected with current Integral Field Unit (IFU) instruments, such as VLT/MUSE \citep[][]{BAA10}.

\section{Numerical simulation}
\label{secsim}

We focus on one high-resolution cosmological zoom simulation taken from the Auriga simulation suite \citep[see][and Grand et al. in preparation for a full description]{GSG16}, performed with the state-of-the-art magneto-hydrodynamical moving-mesh code \textlcsc{AREPO} \citep{Sp10}. The halo was initially selected from a parent dark matter only cosmological simulation of comoving periodic box size 100 Mpc, with the standard $\Lambda$CDM cosmology. The adopted cosmological parameters are $\Omega _m = 0.307$, $\Omega _b = 0.048$, $\Omega _{\Lambda} = 0.693$ and a Hubble constant of $H_0 = 100 h$ km s$^{-1}$ Mpc$^{-1}$, where $h = 0.6777$, taken from \citet{PC13}. At the end-point of this simulation, candidate halos were selected within a narrow mass range interval around $10^{12} \rm M_{\odot}$ which are located at least $1.37$ Mpc from any object more than half the mass of the candidate at $z=0$, in order to select a sample of Milky Way size systems that are relatively isolated. At $z=127$, the resolution of the dark matter particles of this halo is increased and gas added to create the initial conditions of the zoom, which is then evolved to present day. The typical mass of a high resolution dark matter particle is $\sim 3 \times 10^{5}$ $\rm M_{\odot}$, and the baryonic mass resolution is $\sim 4 \times 10^{4}$ $\rm M_{\odot}$, with a maximum physical softening length equal to $369$ pc \citep[{\bff see}][]{GSG16}. 

The simulation includes a comprehensive galaxy formation model \citep[see][for more details]{VGS13,MPS14,GSG16}, which reproduces many of the global properties representative of observed galaxy populations, such as the stellar to halo mass function, cosmic star formation rate density and galaxy morphologies. The simulation discussed in this study is Au 25 presented in \citet{GSG16}, which is a relatively isolated galaxy that displays clear two-armed spiral structure and is therefore an ideal choice to study the nature of spiral arms in cosmological simulations and their effects on disc chemo-dynamics.

% trim[left, lower, right, upper]

\begin{figure*}
\includegraphics[scale=1.4,trim={0 0.1cm 0.5cm 0},clip]{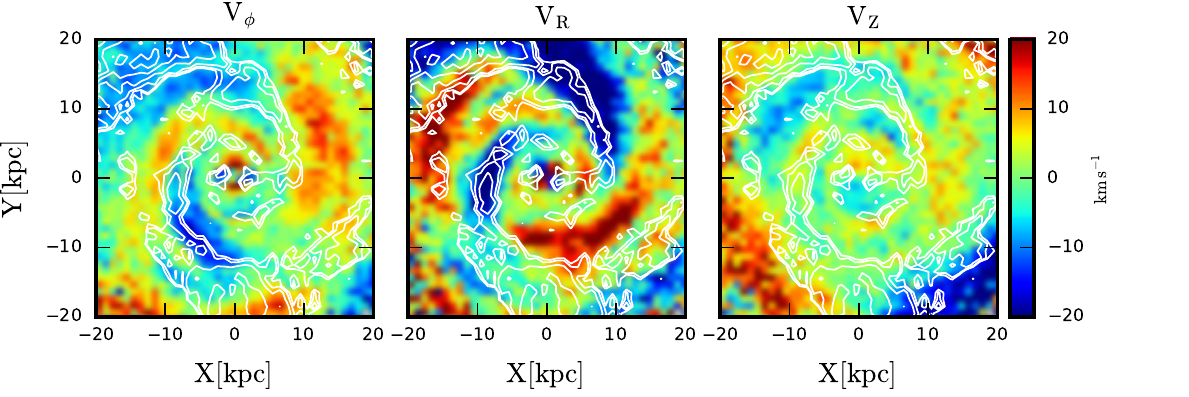}
\caption{Face-on maps of the {\bff azimuthal} (left), radial (middle) and vertical (right) peculiar velocity fields. Positive velocities are in the direction of rotation ({\bff azimuthal}), the galactic anti-centre (radial) and positive vertical heights (vertical). Over-density contours of the mass distribution are indicated by the contours. {\bff The {\bff azimuthal} peculiar velocity field is systematically slower (faster) on the trailing (leading) edge of the spiral, whereas the radial peculiar velocity points outward (inward) on the trailing (leading) edge.} The amplitude of the fluctuations in the vertical peculiar velocity field are lower than the planar velocity fields, and show a less coherent pattern.}
\label{vint}
\end{figure*}

\begin{figure}\centering
\includegraphics[scale=1.75,trim={0 0.75cm 0 0.5cm},clip]{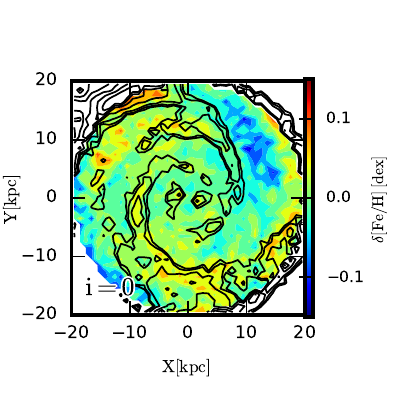}\\ %\vspace{0.1cm}
\includegraphics[scale=1.75,trim={0 0.1cm 0 0.6cm},clip]{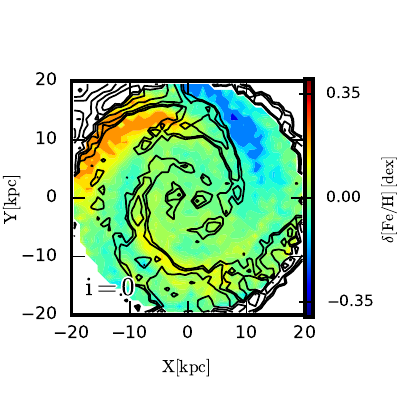} 
\caption{Face-on map (inclination, $i=0$) of the azimuthal residual of the actual metallicity field (top panel) and {\bff that obtained for the case in which the} metallicity field {\bff is} artificially generated 120 Myr before the current time (bottom panel). Over-density contours of the mass distribution are indicated by the contours. At many radii there is an over-density of metal rich (poor) star particles on the trailing (leading) side of the spiral arm.}
\label{metint}
\end{figure}

\section{Results}

Spiral structure is situated in the disc component of galaxies, and it follows that stars that spend much of their orbit in the disc mid-plane are dynamically responsive to such structures. We therefore focus our analysis on young stars (age $< 3$ Gyr) that belong to the thin disc, {\bff which constitute about $27\%$ of the total stellar mass within a radius of 25 kpc. Note that if all star particles were to be considered, the observable features described in this Letter would be weaker, owing to contamination of bulge and halo stars.} We focus on a single snapshot of galaxy Au 25 at a lookback time of $2.67$ Gyr, which we choose because of its late time, quiescent environment and well-formed spiral arms in order to demonstrate clearly the dynamical signatures related to spiral structure, which is the aim of this study. The nature of the spiral arms and their evolution will be studied in a forthcoming paper.

\subsection{Velocity fields}

In the following, we define the azimuthal peculiar velocity, $V_{\phi}$, as the difference between the azimuthal velocity of a star particle and the mean rotation velocity at the particle radius, and define $V_{\phi} > 0$ as faster than mean rotation. The radial and vertical peculiar velocities, $V_R$ and $V_Z$, are defined as the radial and vertical velocities, with $V_R > 0$ toward the galactic anti-centre and $V_Z > 0$ toward the north galactic pole.

In the left panel of Fig.~\ref{vint} we show the face-on map of $V_{\phi}$, with azimuthal over-density contours of the mass distribution, given by $(\Sigma (R,\phi) - \Sigma(R)) / \Sigma (R)$, overlaid in white contours. The spiral structure extends from about $5$ to $15$ kpc, and is accompanied by a well-defined spiral-shaped pattern in the  azimuthal peculiar velocity field: stars rotate locally slower on the trailing side of the spiral arm, whereas they rotate locally faster on the leading side. The middle panel of Fig.~\ref{vint} shows the face-on map of $V_R$. Similarly, this velocity field reveals a spiral shaped pattern in which the spiral arm locus delineates the outward and inward streaming motions that are situated on the trailing and leading sides of the spirals, respectively. The right panel of Fig.~\ref{vint} shows the face-on map of $V_Z$, the fluctuations of which are of a lower amplitude in comparison to the planar velocity fields. We note that there may be indications of vertical modes in this galaxy \citep[such as those shown in][]{GWM15}, which will be investigated in a forthcoming paper (Gomez et al. in preparation).

These patterns in the velocity fields are qualitatively similar to the systematic motions discussed in several recent studies of idealised simulations of isolated discs \citep[e.g.,][]{KHG14,HKG15,GBK15} and in some observational work \citep[e.g.,][]{CRS15}. In the former studies, the systematic motions have been linked to transient, winding spiral density enhancements commonly seen in $N$-body simulations, in which star particles on the trailing or leading side of the spiral maintain their position with respect to the spiral peak, and are continuously torqued to larger or smaller guiding centre radii \citep{GKC11,GKC12,GKC13b}.

\begin{figure}
\includegraphics[scale=1.3,trim={0.2cm 0.1cm 0.4cm 0},clip]{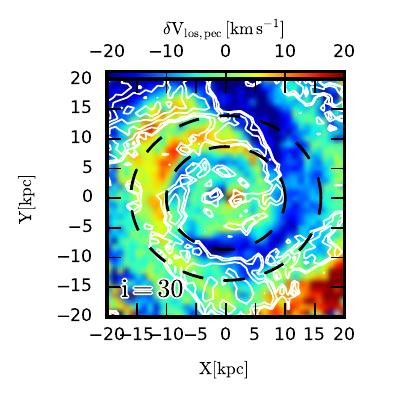} 
\includegraphics[scale=1.3,trim={0.7cm 0.1cm 0.4cm 0},clip]{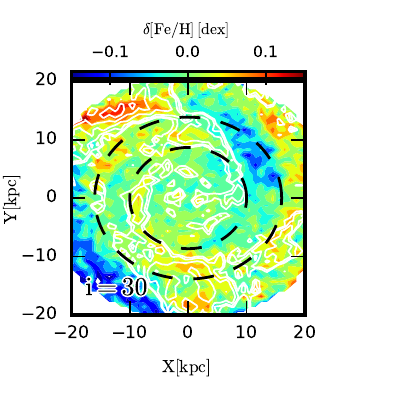} \\
\includegraphics[scale=1.16,trim={0.85cm 0.1cm 0 0.2cm},clip]{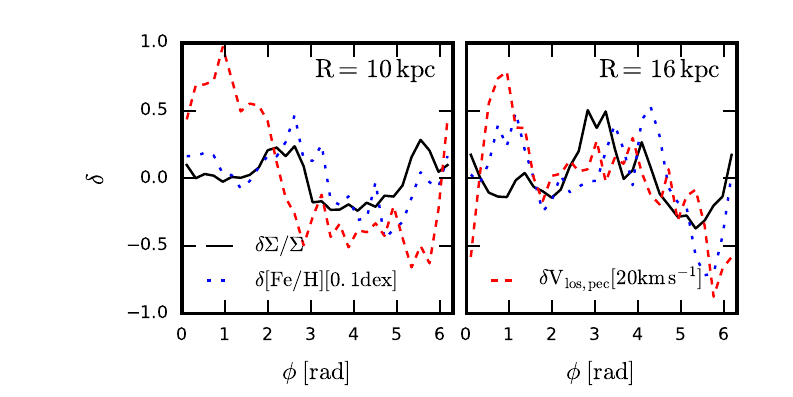}
\caption{\emph{Top:} The peculiar LOS velocity field (left) and residual metallicity (right) of the disc at an inclination of 30 degrees. Positive LOS velocities are directed away from the observer. The dashed lines indicate galacto-centric radii of 10 and 16 kpc. \emph{Bottom:} The azimuthal profiles of the surface density (black solid), residual metallicity (blue dotted) and peculiar LOS velocity (red dashed) at radii of 10 (left) and 16 (right) kpc. $\phi = 0$ corresponds to $X=0$ for $Y > 0$, and increases with counter-clockwise rotation in the disc plane.}
\label{los}
\end{figure}

\subsection{Metal distribution}

The torques applied to stars from spiral arms that generate the systematic streaming motions play a significant role in re-distributing individual stars around the disc, a process referred to as radial migration \citep{SB02,MF10,GKC11}. For a disc with a negative radial metallicity gradient, such as that of the Milky Way \citep[e.g.,][]{BSP13,BRS14,ACS14}, the radial re-distribution of stars can lead to changes in the metal distribution, in particular metal rich (poor) stars from the inner (outer) regions are brought to the outer (inner) regions. Such a radial re-distribution of stars has been shown to broaden the metallicity distribution \citep{MCM13,GKC15}, and is required to explain the large scatter in the age-metallicity relation in the solar neighbourhood \citep{CM01,H08,CSA11}. However, the observation of these trends is possible in the Milky Way only, for which star-by-star measurements are available. %This is currently not possible to achieve for external galaxies.

A more directly observable signature of radial migration along spiral arms may come from azimuthal trends of metallicity at a given radius, in much the same spirit as the velocity trends are found in Fig.~\ref{vint}. To date, the only study of such a signature from a simulation perspective is that of \citet{DiM13}, who studied the azimuthal variation of the metallicity distribution of old stars around a bar and found a pattern characteristic of radial migration around the bar co-rotation radius. A detailed {\bff study} of the azimuthal variations of chemical abundances in APOGEE data, and their relation to Milky Way spiral structure, will be presented in Minchev et al. in preparation \citep[see also][]{BNR14}.

In the top panel of Fig.~\ref{metint}, we show the face-on map of the azimuthal residual metallicity distribution, defined as $\delta \mathrm{[Fe/H]} (R,\phi) =  \mathrm{[Fe/H]} (R,\phi) -  \overline{\mathrm{[Fe/H]}}(R)$. It is clear that in the radial range of spiral structure the metallicity pattern is characterised by an over-density of metal rich stars on the trailing side of the spiral arm, whereas an over-density of metal poor stars are found on the leading side of the spiral. {\bff These features are a consequence of the radial metallicity distribution, which has a radial gradient of $-0.035$ $\rm dex$ $\rm kpc^{-1}$ and metallicity dispersion of about $0.19$ dex at a given radius.} We note that the trend is even more clear in the bottom panel of Fig.~\ref{metint}, in which the radial metallicity distribution is artificially set 120 Myr earlier (about a dynamical time) with a fixed radial gradient of $-0.08$ $\rm dex$ $\rm kpc^{-1}$ and a radially constant metallicity dispersion of $0.05$ dex. This confirms that the dynamics are consistent with large-scale radial migration along the spiral arms, and also that the signatures become more clear for  steeper radial gradients and narrower metallicity dispersions, respectively. This is the first time that such a trend has been shown, and provides a further observational test for the radial migration driven by spiral arms.

\subsection{Observing the line-of-sight signatures}

The features of the peculiar velocity and residual metallicity fields shown above can be directly observed in external galaxies with {\bff IFU} instruments, such as VLT/MUSE. To demonstrate how the peculiar velocity field and metallicity distribution are mapped to the line-of-sight (LOS) velocity field\footnote{ We note that the contribution from the vertical peculiar velocity to the LOS velocity within this radial range is minor and does not affect the trends discussed in this paper.}, $\rm \delta V_{los,pec}$, and LOS metallicity distribution, we {\bff set the disc to} an inclination of $30$ degrees{\bff , which is enough inclined to observe planar peculiar velocities while maintaining a clear view of the spiral structure. The projected} LOS peculiar velocity map and residual metallicity map {\bff are shown} in the top panels of Fig.~\ref{los}. For $X<0$, positive $V_{\phi}$ is mapped to positive LOS velocities (away from the observer), and to negative LOS velocities (toward the observer) for $X>0$. Positive $V_R$ is mapped to positive LOS velocities for $Y>0$, and negative LOS velocities for $Y<0$.

To quantify the fluctuations in both fields, we show in the bottom panels of Fig.~\ref{los} the azimuthal profiles of the residuals of the mass surface density, metallicity and LOS peculiar velocity fields at two different radii. {\bff We define $\phi = 0$ for $X=0$ and $Y>0$, which increases counter-clockwise in the disc plane.} The patterns in the LOS peculiar velocity, $\rm \delta V_{los,pec}$, depend on the location of the spiral arms, because the direction of the LOS projection of the radial and {\bff azimuthal} peculiar velocities flips around $Y=0$ and $X=0$, respectively. At $R=10$ kpc, the spiral arm at $\phi \sim 5.5$ rad is located at $Y > 0$. Both the {\bff leading ($\phi < 5.5$ rad) and trailing side ($\phi > 5.5$ rad)}, have positive $X$ and $Y$ coordinates, where {\bff azimuthal}ly fast (slow) and radially inward (outward) peculiar velocities on the leading (trailing) edge give rise to negative (positive) $\rm \delta V_{los,pec}$. Clockwise (decreasing $\phi$) of this spiral arm the $\rm \delta V_{los,pec}$ remains negative because the radially positive motion on the trailing edge of the next spiral ($\phi \sim 3.0$ rad) for $Y<0$ is directed toward the observer. On the leading edge of this spiral arm ($\phi \sim 1.5$ rad) the {\bff large peculiar {\bff azimuthal}} velocity leads to a positive $\rm \delta V_{los,pec}$.  A similar trend is present at $R=16$ kpc, though there is {\bff a small additional peak at $\phi \sim 4.0$ owing to} the phase shift in spiral arm position with respect to $R=10$ kpc {\bff that causes negative peculiar {\bff azimuthal} velocities on the trailing side of the spiral to contribute to positive $\rm \delta V_{los,pec}$ for $X>0$ and $Y<0$}. The {\bff semi}amplitude of $\rm \delta V_{los,pec}$ is about $10-15$ $\rm km \, s^{-1}$. For the metallicity residuals, the {\bff semi}amplitude is $\sim 0.05$ dex, which yields a total variation of  $\sim 0.1$ dex. The {\bff semi}amplitudes of the LOS peculiar velocity can be increased up to $\sim 20$ $\rm km \, s^{-1}$ for inclinations up to $\sim 60$ degrees, however the spatial resolution is lower. The magnitudes of these fluctuations should be large enough to be detected with {\bff IFU} observations of nearby late-type galaxies.

\section{Conclusions}

In this study we have analysed signatures in the peculiar velocity and residual metallicity fields linked to the dynamical influence of spiral arms, in one of the high resolution, fully cosmological zoom simulations from the Auriga suite \citep{GSG16}. We have demonstrated that the peculiar {\bff azimuthal} velocity is locally slower (faster) on the trailing (leading) edge of the spiral arm. Similarly, the peculiar radial velocity is directed radially outward (inward) on the trailing (leading) edge of the spiral arm, which in combination with the {\bff azimuthal} velocity creates a systematic streaming motion along the spiral arm. This represents the first confirmation of systematic radial migration around spiral arms in a fully cosmological zoom simulation, which is in agreement with idealised $N$-body simulations of isolated discs that show transient, winding spiral arms. 

In addition, we show for the first time that the radial migration caused by the spiral arms leads to azimuthal variations of the metallicity distribution: at a given radius, star particles that originated from interior regions of the disc are metal rich with respect to the azimuthal mean metallicity at that radius, because of the negative metallicity gradient commonly observed in disc galaxies. As indicated by the systematic streaming motions, the metal rich particles are transported outward along the trailing edge of the spiral. Similarly, the metal poor star particles that originate in the outer disc regions are transported radially inwards along the leading edge of the spiral. The result of these motions is a residual metallicity pattern in azimuth which is systematically more metal rich (poor) along the trailing (leading) edge of the spiral arm at many radii. 

Finally, we have shown that the azimuthal variations of the peculiar LOS velocity and metallicity maps in a disc inclined at an angle of 30 degrees are about 20-30 $\rm km \, s^{-1}$ and 0.1 dex, respectively. These variations can be detectable in nearby late-type spiral galaxies with IFU instruments such as VLT/MUSE. 

We note that a systematic difference in azimuthal velocity across a density wave-like spiral arm has been suggested by \citet{MQ08,PNK15}. However, these motions would depend on the location of the resonance points and are not expected to drive a metallicity variation around the spiral arm, because they are not {\bff associated with radial} migration. The results of this Letter therefore represent a prediction of the dynamical influence of spiral arms on stars that differs qualitatively from that of density wave-like spirals \citep[see also][]{MFS16}, and have the potential to test spiral arm formation mechanisms.

This Letter is an encouraging first step toward making new testable predictions of the systematic radial migration driven by spiral arms in fully cosmological simulations, which we have demonstrated are sufficiently advanced to study spiral arms. The exact values of the difference in the LOS velocities and residual azimuthal metallicity likely depend on the amplitude and pitch angle of spiral arms, the metallicity gradient, and the other properties of the thin disc. In future work, we will study a variety of spiral arm features using both {\bff cosmological and idealised simulations.}

\section*{acknowledgements}
{\bff We thank the anonymous referee for a constructive report.} RG and VS acknowledge support by the DFG Research Centre SFB-881 `The Milky Way System' through project A1. DJRC acknowledges STFC studentship ST/K501979/1. This work has also been supported by the European Research Council under ERC-StG grant EXAGAL- 308037. Part of the simulations of this paper used the SuperMUC system at the Leibniz Computing Centre, Garching, under the project PR85JE of the Gauss Centre for Supercomputing. This work used the DiRAC Data Centric system at Durham University, operated by the Institute for Computational Cosmology on behalf of the STFC DiRAC HPC Facility `www.dirac.ac.uk'. This equipment was funded by BIS National E-infrastructure capital grant ST/K00042X/1, STFC capital grant ST/H008519/1, and STFC DiRAC Operations grant ST/K003267/1 and Durham University. DiRAC is part of the National E-Infrastructure.

\bibliographystyle{mnras}
\bibliography{GG3d3R1.bbl}

\begin{thebibliography}{}
\makeatletter
\relax
\def\mn@urlcharsother{\let\do\@makeother \do\$\do\&\do\#\do\^\do\_\do\%\do\~}
\def\mn@doi{\begingroup\mn@urlcharsother \@ifnextchar [ {\mn@doi@}
  {\mn@doi@[]}}
\def\mn@doi@[#1]#2{\def\@tempa{#1}\ifx\@tempa\@empty \href
  {http://dx.doi.org/#2} {doi:#2}\else \href {http://dx.doi.org/#2} {#1}\fi
  \endgroup}
\def\mn@eprint#1#2{\mn@eprint@#1:#2::\@nil}
\def\mn@eprint@arXiv#1{\href {http://arxiv.org/abs/#1} {{\tt arXiv:#1}}}
\def\mn@eprint@dblp#1{\href {http://dblp.uni-trier.de/rec/bibtex/#1.xml}
  {dblp:#1}}
\def\mn@eprint@#1:#2:#3:#4\@nil{\def\@tempa {#1}\def\@tempb {#2}\def\@tempc
  {#3}\ifx \@tempc \@empty \let \@tempc \@tempb \let \@tempb \@tempa \fi \ifx
  \@tempb \@empty \def\@tempb {arXiv}\fi \@ifundefined
  {mn@eprint@\@tempb}{\@tempb:\@tempc}{\expandafter \expandafter \csname
  mn@eprint@\@tempb\endcsname \expandafter{\@tempc}}}

\bibitem[\protect\citeauthoryear{{Anders} et~al.}{{Anders}
  et~al.}{2014}]{ACS14}
{Anders} F.,  et~al., 2014, \mn@doi [\aap] {10.1051/0004-6361/201323038}, \href
  {http://adsabs.harvard.edu/abs/2014A%26A...564A.115A} {564, A115}

\bibitem[\protect\citeauthoryear{{Baba}, {Saitoh}  \& {Wada}}{{Baba}
  et~al.}{2013}]{BSW12}
{Baba} J.,  {Saitoh} T.~R.,   {Wada} K.,  2013, \mn@doi [\apj]
  {10.1088/0004-637X/763/1/46}, \href
  {http://adsabs.harvard.edu/abs/2013ApJ...763...46B} {763, 46}

\bibitem[\protect\citeauthoryear{{Bacon} et~al.}{{Bacon} et~al.}{2010}]{BAA10}
{Bacon} R.,  et~al., 2010, in Society of Photo-Optical Instrumentation
  Engineers (SPIE) Conference Series. p.~8, \mn@doi{10.1117/12.856027}

\bibitem[\protect\citeauthoryear{{Bergemann} et~al.}{{Bergemann}
  et~al.}{2014}]{BRS14}
{Bergemann} M.,  et~al., 2014, \mn@doi [\aap] {10.1051/0004-6361/201423456},
  \href {http://adsabs.harvard.edu/abs/2014A%26A...565A..89B} {565, A89}

\bibitem[\protect\citeauthoryear{{Boeche} et~al.}{{Boeche}
  et~al.}{2013}]{BSP13}
{Boeche} C.,  et~al., 2013, \mn@doi [\aap] {10.1051/0004-6361/201322085}, \href
  {http://adsabs.harvard.edu/abs/2013A%26A...559A..59B} {559, A59}

\bibitem[\protect\citeauthoryear{{Bovy} et~al.}{{Bovy} et~al.}{2014}]{BNR14}
{Bovy} J.,  et~al., 2014, \mn@doi [\apj] {10.1088/0004-637X/790/2/127}, \href
  {http://adsabs.harvard.edu/abs/2014ApJ...790..127B} {790, 127}

\bibitem[\protect\citeauthoryear{{Casagrande}, {Sch{\"o}nrich}, {Asplund},
  {Cassisi}, {Ram{\'{\i}}rez}, {Mel{\'e}ndez}, {Bensby}  \&
  {Feltzing}}{{Casagrande} et~al.}{2011}]{CSA11}
{Casagrande} L.,  {Sch{\"o}nrich} R.,  {Asplund} M.,  {Cassisi} S.,
  {Ram{\'{\i}}rez} I.,  {Mel{\'e}ndez} J.,  {Bensby} T.,   {Feltzing} S.,
  2011, \mn@doi [\aap] {10.1051/0004-6361/201016276}, \href
  {http://adsabs.harvard.edu/abs/2011A%26A...530A.138C} {530, A138}

\bibitem[\protect\citeauthoryear{{Chemin}, {Renaud}  \& {Soubiran}}{{Chemin}
  et~al.}{2015}]{CRS15}
{Chemin} L.,  {Renaud} F.,   {Soubiran} C.,  2015, \mn@doi [\aap]
  {10.1051/0004-6361/201526040}, \href
  {http://adsabs.harvard.edu/abs/2015A%26A...578A..14C} {578, A14}

\bibitem[\protect\citeauthoryear{{Chiappini}, {Matteucci}  \&
  {Romano}}{{Chiappini} et~al.}{2001}]{CM01}
{Chiappini} C.,  {Matteucci} F.,   {Romano} D.,  2001, \mn@doi [\apj]
  {10.1086/321427}, \href {http://adsabs.harvard.edu/abs/2001ApJ...554.1044C}
  {554, 1044}

\bibitem[\protect\citeauthoryear{{Di Matteo}, {Haywood}, {Combes}, {Semelin}
  \& {Snaith}}{{Di Matteo} et~al.}{2013}]{DiM13}
{Di Matteo} P.,  {Haywood} M.,  {Combes} F.,  {Semelin} B.,   {Snaith} O.~N.,
  2013, \mn@doi [\aap] {10.1051/0004-6361/201220539}, \href
  {http://adsabs.harvard.edu/abs/2013A%26A...553A.102D} {553, A102}

\bibitem[\protect\citeauthoryear{{Ferreras}, {Cropper}, {Kawata}, {Page}  \&
  {Hoversten}}{{Ferreras} et~al.}{2012}]{FCK12}
{Ferreras} I.,  {Cropper} M.,  {Kawata} D.,  {Page} M.,   {Hoversten} E.~A.,
  2012, \mn@doi [\mnras] {10.1111/j.1365-2966.2012.21017.x}, \href
  {http://adsabs.harvard.edu/abs/2012MNRAS.424.1636F} {424, 1636}

\bibitem[\protect\citeauthoryear{{Foyle}, {Rix}, {Walter}  \& {Leroy}}{{Foyle}
  et~al.}{2010}]{FR10}
{Foyle} K.,  {Rix} H.-W.,  {Walter} F.,   {Leroy} A.~K.,  2010, \mn@doi [\apj]
  {10.1088/0004-637X/725/1/534}, \href
  {http://adsabs.harvard.edu/abs/2010ApJ...725..534F} {725, 534}

\bibitem[\protect\citeauthoryear{{G{\'o}mez}, {White}, {Marinacci}, {Slater},
  {Grand}, {Springel}  \& {Pakmor}}{{G{\'o}mez} et~al.}{2016}]{GWM15}
{G{\'o}mez} F.~A.,  {White} S.~D.~M.,  {Marinacci} F.,  {Slater} C.~T.,
  {Grand} R.~J.~J.,  {Springel} V.,   {Pakmor} R.,  2016, \mn@doi [\mnras]
  {10.1093/mnras/stv2786}, \href
  {http://adsabs.harvard.edu/abs/2016MNRAS.456.2779G} {456, 2779}

\bibitem[\protect\citeauthoryear{{Grand}, {Kawata}  \& {Cropper}}{{Grand}
  et~al.}{2012a}]{GKC11}
{Grand} R.~J.~J.,  {Kawata} D.,   {Cropper} M.,  2012a, \mn@doi [\mnras]
  {10.1111/j.1365-2966.2012.20411.x}, \href
  {http://adsabs.harvard.edu/abs/2012MNRAS.421.1529G} {421, 1529}

\bibitem[\protect\citeauthoryear{{Grand}, {Kawata}  \& {Cropper}}{{Grand}
  et~al.}{2012b}]{GKC12}
{Grand} R.~J.~J.,  {Kawata} D.,   {Cropper} M.,  2012b, \mn@doi [\mnras]
  {10.1111/j.1365-2966.2012.21733.x}, \href
  {http://adsabs.harvard.edu/abs/2012MNRAS.426..167G} {426, 167}

\bibitem[\protect\citeauthoryear{{Grand}, {Kawata}  \& {Cropper}}{{Grand}
  et~al.}{2013}]{GKC13}
{Grand} R.~J.~J.,  {Kawata} D.,   {Cropper} M.,  2013, \mn@doi [\aap]
  {10.1051/0004-6361/201321308}, \href
  {http://adsabs.harvard.edu/abs/2013A%26A...553A..77G} {553, A77}

\bibitem[\protect\citeauthoryear{{Grand}, {Kawata}  \& {Cropper}}{{Grand}
  et~al.}{2014}]{GKC13b}
{Grand} R.~J.~J.,  {Kawata} D.,   {Cropper} M.,  2014, \mn@doi [\mnras]
  {10.1093/mnras/stt2483}, \href
  {http://adsabs.harvard.edu/abs/2014MNRAS.439..623G} {439, 623}

\bibitem[\protect\citeauthoryear{{Grand}, {Kawata}  \& {Cropper}}{{Grand}
  et~al.}{2015a}]{GKC15}
{Grand} R.~J.~J.,  {Kawata} D.,   {Cropper} M.,  2015a, \mn@doi [\mnras]
  {10.1093/mnras/stv016}, \href
  {http://adsabs.harvard.edu/abs/2015MNRAS.447.4018G} {447, 4018}

\bibitem[\protect\citeauthoryear{{Grand}, {Bovy}, {Kawata}, {Hunt}, {Famaey},
  {Siebert}, {Monari}  \& {Cropper}}{{Grand} et~al.}{2015b}]{GBK15}
{Grand} R.~J.~J.,  {Bovy} J.,  {Kawata} D.,  {Hunt} J.~A.~S.,  {Famaey} B.,
  {Siebert} A.,  {Monari} G.,   {Cropper} M.,  2015b, \mn@doi [\mnras]
  {10.1093/mnras/stv1785}, \href
  {http://adsabs.harvard.edu/abs/2015MNRAS.453.1867G} {453, 1867}

\bibitem[\protect\citeauthoryear{{Grand}, {Springel}, {G{\'o}mez}, {Marinacci},
  {Pakmor}, {Campbell}  \& {Jenkins}}{{Grand} et~al.}{2016}]{GSG16}
{Grand} R.~J.~J.,  {Springel} V.,  {G{\'o}mez} F.~A.,  {Marinacci} F.,
  {Pakmor} R.,  {Campbell} D.~J.~R.,   {Jenkins} A.,  2016, \mn@doi [\mnras]
  {10.1093/mnras/stw601}, \href
  {http://adsabs.harvard.edu/abs/2016MNRAS.tmp..395G} {}

\bibitem[\protect\citeauthoryear{{Haywood}}{{Haywood}}{2008}]{H08}
{Haywood} M.,  2008, \mn@doi [\mnras] {10.1111/j.1365-2966.2008.13395.x}, \href
  {http://adsabs.harvard.edu/abs/2008MNRAS.388.1175H} {388, 1175}

\bibitem[\protect\citeauthoryear{{Hunt}, {Kawata}, {Grand}, {Minchev},
  {Pasetto}  \& {Cropper}}{{Hunt} et~al.}{2015}]{HKG15}
{Hunt} J.~A.~S.,  {Kawata} D.,  {Grand} R.~J.~J.,  {Minchev} I.,  {Pasetto} S.,
    {Cropper} M.,  2015, \mn@doi [\mnras] {10.1093/mnras/stv765}, \href
  {http://adsabs.harvard.edu/abs/2015MNRAS.450.2132H} {450, 2132}

\bibitem[\protect\citeauthoryear{{Julian} \& {Toomre}}{{Julian} \&
  {Toomre}}{1966}]{JT66}
{Julian} W.~H.,  {Toomre} A.,  1966, \mn@doi [\apj] {10.1086/148957}, \href
  {http://adsabs.harvard.edu/abs/1966ApJ...146..810J} {146, 810}

\bibitem[\protect\citeauthoryear{{Kawata}, {Hunt}, {Grand}, {Pasetto}  \&
  {Cropper}}{{Kawata} et~al.}{2014}]{KHG14}
{Kawata} D.,  {Hunt} J.~A.~S.,  {Grand} R.~J.~J.,  {Pasetto} S.,   {Cropper}
  M.,  2014, \mn@doi [\mnras] {10.1093/mnras/stu1292}, \href
  {http://adsabs.harvard.edu/abs/2014MNRAS.443.2757K} {443, 2757}

\bibitem[\protect\citeauthoryear{{Lin} \& {Shu}}{{Lin} \& {Shu}}{1964}]{LS64}
{Lin} C.~C.,  {Shu} F.~H.,  1964, \mn@doi [\apj] {10.1086/147955}, \href
  {http://adsabs.harvard.edu/abs/1964ApJ...140..646L} {140, 646}

\bibitem[\protect\citeauthoryear{{Marinacci}, {Pakmor}  \&
  {Springel}}{{Marinacci} et~al.}{2014}]{MPS14}
{Marinacci} F.,  {Pakmor} R.,   {Springel} V.,  2014, \mn@doi [\mnras]
  {10.1093/mnras/stt2003}, \href
  {http://adsabs.harvard.edu/abs/2014MNRAS.437.1750M} {437, 1750}

\bibitem[\protect\citeauthoryear{{Michikoshi} \& {Kokubo}}{{Michikoshi} \&
  {Kokubo}}{2014}]{MK14}
{Michikoshi} S.,  {Kokubo} E.,  2014, \mn@doi [\apj]
  {10.1088/0004-637X/787/2/174}, \href
  {http://adsabs.harvard.edu/abs/2014ApJ...787..174M} {787, 174}

\bibitem[\protect\citeauthoryear{{Minchev} \& {Famaey}}{{Minchev} \&
  {Famaey}}{2010}]{MF10}
{Minchev} I.,  {Famaey} B.,  2010, \mn@doi [\apj]
  {10.1088/0004-637X/722/1/112}, \href
  {http://adsabs.harvard.edu/abs/2010ApJ...722..112M} {722, 112}

\bibitem[\protect\citeauthoryear{{Minchev} \& {Quillen}}{{Minchev} \&
  {Quillen}}{2008}]{MQ08}
{Minchev} I.,  {Quillen} A.~C.,  2008, \mn@doi [\mnras]
  {10.1111/j.1365-2966.2008.13134.x}, \href
  {http://adsabs.harvard.edu/abs/2008MNRAS.386.1579M} {386, 1579}

\bibitem[\protect\citeauthoryear{{Minchev}, {Chiappini}  \& {Martig}}{{Minchev}
  et~al.}{2013}]{MCM13}
{Minchev} I.,  {Chiappini} C.,   {Martig} M.,  2013, \mn@doi [\aap]
  {10.1051/0004-6361/201220189}, \href
  {http://adsabs.harvard.edu/abs/2013A%26A...558A...9M} {558, A9}

\bibitem[\protect\citeauthoryear{{Monari}, {Famaey}  \& {Siebert}}{{Monari}
  et~al.}{2016}]{MFS16}
{Monari} G.,  {Famaey} B.,   {Siebert} A.,  2016, \mn@doi [\mnras]
  {10.1093/mnras/stw171}, \href
  {http://adsabs.harvard.edu/abs/2016MNRAS.457.2569M} {457, 2569}

\bibitem[\protect\citeauthoryear{{Pasetto}, {Natale}, {Kawata}, {Chiosi}  \&
  {Hunt}}{{Pasetto} et~al.}{2015}]{PNK15}
{Pasetto} S.,  {Natale} G.,  {Kawata} D.,  {Chiosi} C.,   {Hunt} J.~A.~S.,
  2015, preprint, \href {http://adsabs.harvard.edu/abs/2015arXiv151205367P} {}
  (\mn@eprint {arXiv} {1512.05367})

\bibitem[\protect\citeauthoryear{{Planck Collaboration} et~al.,}{{Planck
  Collaboration} et~al.}{2014}]{PC13}
{Planck Collaboration} et~al., 2014, \mn@doi [\aap]
  {10.1051/0004-6361/201321591}, \href
  {http://adsabs.harvard.edu/abs/2014A%26A...571A..16P} {571, A16}

\bibitem[\protect\citeauthoryear{{Rix} \& {Zaritsky}}{{Rix} \&
  {Zaritsky}}{1995}]{RZ95}
{Rix} H.-W.,  {Zaritsky} D.,  1995, \mn@doi [\apj] {10.1086/175858}, \href
  {http://adsabs.harvard.edu/abs/1995ApJ...447...82R} {447, 82}

\bibitem[\protect\citeauthoryear{{Sellwood}}{{Sellwood}}{2011}]{Se11}
{Sellwood} J.~A.,  2011, \mn@doi [\mnras] {10.1111/j.1365-2966.2010.17545.x},
  \href {http://adsabs.harvard.edu/abs/2011MNRAS.410.1637S} {410, 1637}

\bibitem[\protect\citeauthoryear{{Sellwood} \& {Binney}}{{Sellwood} \&
  {Binney}}{2002}]{SB02}
{Sellwood} J.~A.,  {Binney} J.~J.,  2002, \mn@doi [\mnras]
  {10.1046/j.1365-8711.2002.05806.x}, \href
  {http://adsabs.harvard.edu/abs/2002MNRAS.336..785S} {336, 785}

\bibitem[\protect\citeauthoryear{{Springel}}{{Springel}}{2010}]{Sp10}
{Springel} V.,  2010, \mn@doi [\mnras] {10.1111/j.1365-2966.2009.15715.x},
  \href {http://adsabs.harvard.edu/abs/2010MNRAS.401..791S} {401, 791}

\bibitem[\protect\citeauthoryear{{Tremaine} \& {Weinberg}}{{Tremaine} \&
  {Weinberg}}{1984}]{TW84}
{Tremaine} S.,  {Weinberg} M.~D.,  1984, \mn@doi [\apjl] {10.1086/184292},
  \href {http://adsabs.harvard.edu/abs/1984ApJ...282L...5T} {282, L5}

\bibitem[\protect\citeauthoryear{{Vogelsberger}, {Genel}, {Sijacki}, {Torrey},
  {Springel}  \& {Hernquist}}{{Vogelsberger} et~al.}{2013}]{VGS13}
{Vogelsberger} M.,  {Genel} S.,  {Sijacki} D.,  {Torrey} P.,  {Springel} V.,
  {Hernquist} L.,  2013, \mn@doi [\mnras] {10.1093/mnras/stt1789}, \href
  {http://adsabs.harvard.edu/abs/2013MNRAS.436.3031V} {436, 3031}

\makeatother
\end{thebibliography}

\end{document}